\documentstyle[12pt,twoside]{article}
\begin{document}
\begin{center}

{\Large\bf 
MASS POWER SPECTRUM IN A UNIVERSE\\[5PT]
DOMINATED BY THE CHAPLYGIN GAS\\[5PT]}
\medskip
 
{\bf
J. C. Fabris\footnote{e-mail: fabris@cce.ufes.br},
S.V.B. Gon\c{c}alves\footnote{e-mail: sergio@cce.ufes.br} and P.E. de Souza\footnote{e-mail: patricia.ilus@bol.com.br}}  \medskip

Departamento de F\'{\i}sica, Universidade Federal do Esp\'{\i}rito Santo, 
CEP29060-900, Vit\'oria, Esp\'{\i}rito Santo, Brazil \medskip

\end{center}
 
\begin{abstract}
The mass power spectrum for a Universe dominated by the Chaplygin gas is evaluated numerically from
scales of the order of the Hubble horizon to $100\;Mpc$. The results are compared with a pure
baryonic Universe and a cosmological constant model. In all three cases, the spectrum increases
with $k$, the wavenumber of the perturbations. The slope of the spectrum is higher for the baryonic model and smaller for the cosmological constant model, the Chaplygin gas interpolating these
two models. The results are analyzed in terms of the sound velocity of the Chaplygin gas and the
moment the Universe begins to accelerate.
\vspace{0.7cm}
\newline
PACS number(s): 98.80.Bp, 98.65.Dx
\end{abstract}

\section{Introduction}

Since the observations of the supernova type Ia have indicate that the Universe must
be in an accelerated expansion today \cite{riess,perlmutter} the nature of the fluid responsible
for this inflationary behaviour has been object of many studies. In order this accelerated expansion
to take place today, the Universe must be dominated by a fluid of negative pressure. Moreover,
this fluid must remain a smooth component of the matter content of the Universe, since it does not
appear in the dynamics of clusters of galaxies, for example. The most natural candidate for this
"dark energy" component is a cosmological constant \cite{carroll}. However, there are two main problems
with the idea that a cosmological term must be the dominant component of the Universe today:
First, in spite of the fact that quantum field theory predicts a cosmological constant, the observational
value would be $120$ orders of magnitude smaller than the predicted value; second, it is a quite special fact that
the energy density associated to a cosmological constant have a value near to the density of other types
of matter exactly today, a fact that is usually called "cosmic coincidence". A very popular way of coping
with these problems is through the introduction of a self-interacting scalar field, called "quintessence"
\cite{steinhardt,sahni}. For some kind of potential terms, which have their justification in fundamental theories
like supergravity, there are the so-called tracking solutions: The self-interacting scalar field evolves in such a way
that it approaches a cosmological constant behaviour exactly today \cite{brax}. But the great variety of possible potential
terms with these properties, and the necessity of fine tuning some parameters in order to generate them,
makes quintessence a promising scenario, but without any definitive model until now.
\par
In a previous work \cite{patricia}, we have considered the possibility that the Universe today is dominated
by another kind of fluid, called Chaplygin gas. The Chaplygin gas is characterized by an equation of state
\begin{equation}
\label{eq-est}
p = - \frac{A}{\rho} \quad , \quad A = \mbox{constant} \quad .
\end{equation}
Hence, it exhibits negative pressure, as required to explain the acceleration of the Universe
today. However, the fact that it depends on the inverse of the density gives it some special properties.
In particular, the conservation of the energy-momentum tensor implies that the Chaplygin gas density depends
on the scale factor of the Universe $a$ as
\begin{equation}
\rho = \sqrt{A + \frac{B}{a^6}} \quad ,
\end{equation}
where $B$ is an integration constant.
This relation, first presented in \cite{pasquier} where some cosmological consequences were explored,
has some interesting properties: For small values of the scale factor, the Chaplygin gas exhibits the
same behaviour as a pressurelless fluid; for large value of the scale factor, it approaches the equation
of state of a cosmological constant. Hence, in some sense it reproduces the quintessence scenario, interpolating a matter
dominated Universe and a cosmological constant dominated Universe, but in quite different way, since no appeal
is in principle made for a self-interacting scalar field.
\par
The Chaplygin gas has been first identified in the study of adiabatic fluids \cite{chaplygin}. More recently,
it has received an interesting motivation, connected with string theories \cite{hoppe,ogawa}. Considering
a $D$-brane in a $D+2$-dimensional space-time and employing the light-cone parametrization, the Nambu-Goto
action reduces itself to the action of a newtonian fluid which obeys the equation of state (\ref{eq-est}).
Even if this reduced action is that of a newtonian fluid, the related group of symmetry has the same dimension
as the Poincar\'e group, revealing that the relativistic origin is somehow hidden in that equation of state
\cite{jackiw}.
In this sense, when we speak about the Chaplygin gas, we can be speaking about a gas of $D$-branes in a
$D+2$-dimensional space-time.
\par
The interest of this equation of state for cosmology can be summarized by evoking two main properties of
the Chaplygin gas in an expanding Universe. First, as it has already been said,
it behaves initially as pressurelles fluid, and later
as a cosmological constant. In this sense, it plays the same role as the scalar field in
the quintessence program. Second, the sound velocity of the Chaplygin gas is positive, in spite of the
fact that its pressure is negative. This a very important property, since, as it has already been shown
in \cite{jerome}, fluids with negative pressure obeying a barotropic equation of state suffer from
instabilities at small scales due to an imaginary sound velocity.
\par
In \cite{patricia}, the evolution of density perturbation in a Universe dominated by the Chaplygin gas
has been analyzed exploiting the fact that the Chaplygin gas obeys the newtonian equations of hydrodynamic.
It has been shown that the predicted scenario is compatible with the formation of local structures, and
that the perturbations in the Chaplygin gas tends to vanish asymptotically, as we should expect due to the
asymptotical behaviour similar to a cosmological constant. In the present paper, we intend to go a little
further. Employing the relativistic equation for the description of the evolution of the Universe, considering
a matter content composed of cold dark matter, baryons, radiation and the Chaplygin gas, we study the evolution of
the power spectrum of the clustered matter. Starting with a given primordial power spectrum,
which is very close to the Harrison-Zeldovich spectrum, at the end
of the radiative phase, we show that this power spectrum evolves to an increasing one at moments near
today. In order to track the specificity of role played by the Chaplygin gas, we perform the same analysis
to the case where the Chaplygin gas is replaced by a cosmological constant and to the case of a Universe
composed of radiation plus
pressurelles matter only. The results indicate that the spectrum predicted by the Chaplygin gas model interpolates
between that of a baryonic Universe (which predicts a strong increasing spectrum today for the scales analyzed here and the primordial spectrum employed)
and a cosmological constant model
(which predicts a moderately spectrum today for the same scales) as the sound velocity associated with the Chaplygin gas
goes from zero ($A = 0$) to the velocity of light.
\par
The model developed here has some important simplifications. We consider, for example, the neutrino and photon
contribution to the radiative fluid, whithout taking into account the different moments of decoupling of
these different components. Moreover, we do not take into account that after the beginning of the material
phase, photons remain coupled to the baryonic matter until the decoupling moment. In the same way we
do not take into account that the cold dark matter fluid decouples from radiation before the baryonic fluid.
We restrict our
analysis to scales that goes from $100\;Mpc$ until the Hubble horizon at the present moment, and we do not
consider the neutrino free streaming and photon diffusion process that occurs when the perturbations are inside the
horizon. All these simplications limit the possibility of comparing in detail the obtained results with the recent surveys
of clustered matter. But we emphasize that the goal of the present paper is to identify some main features
of the Chaplygin gas concerning the structure formation problem, and in view of this goal those simplifications
(even if very strong from the observational point of view)
do not spoil our main conclusions. However, the results are qualitative in agreement with more precise and
realistic calculations.
\par
This paper is organized as follows. In next section, we describe the cosmological model with the Chaplygin gas,
comparing it with a baryonic Universe and a cosmological constant model.
In section 3, we settle out the perturbative equations which are analyzed
numerically. The power spectrum is evaluated for each of the models described before. In section 4, we present our
conclusions.

\section{The background model and the cosmological parameters}

Let us consider a quite general model consisting of a pressurelles fluid, radiation, a cosmological constant term
and the Chaplygin gas. If we take a Friedmann-Robertson-Walker metric, which describes an homogeneous and isotropic Universe, the Einstein's equations reduce to
\begin{eqnarray}
\biggr(\frac{\dot a}{a}\biggl)^2 + \frac{k}{a^2} &=& \frac{8\pi G}{3}(\rho_m + \rho_r + \rho_\Lambda + \rho_c) \quad , \\
\frac{\ddot a}{a} + 2\biggr(\frac{\dot a}{a}\biggl)^2 + \frac{k}{a^2} &=& - 8\pi G(p_m + p_r - \rho_\Lambda + p_c) \quad , 
\end{eqnarray}
where $a$ is the scale factor of the Universe, $k$ is the curvature of the spatial section,
and $\rho_m$, $\rho_r$ and $\rho_c$ are the density for the pressurelles fluid, radiation and the Chaplygin gas,
$p_m = 0$, $p_r = \rho_r/3$ and $p_c = - A/\rho_c$ being their corresponding pressures.
$\rho_\Lambda$ is the density associated to the cosmological term. In what follows we will be interested in three cases:
$\rho_\Lambda = 0$, the Chaplygin gas model; $\rho_c$ = 0, the cosmological constant model; $\rho_\Lambda = \rho_c = 0$, the baryonic model.
\par
The curvature of the spatial section will be set equal zero. This simplification is justified by the fact that
the data concerning the anisotropy of the cosmic microwave background radiation favor a flat, or almost flat,
Universe. In fact, the parameter associated with the curvature of the spatial section of the Universe is
defined as $\Omega_k = 1 - \Omega_T$, where $\Omega_T$ is the total observed density of the Universe divided by
the critical density. The position of the first Doppler peak of the spectrum of the anisotropy of the cosmic
microwave background radiation is directly connected to $\Omega_T$. The recent results from the
BOOMERANG and MAXIMA measurements of CMB spectrum give $\Omega_k = 0 \pm 0.06$ \cite{charles}. Hence, we can fix
$k = 0$ without to oversimplify our model. Moreover, an inflationary phase in the primordial Universe
predicts (except for very special cases) $k = 0$.
\par
The different fluids are assumed to interact through the geometry only. Hence, the energy-momentum tensor
for each component conserve separately,
\begin{equation}
{T^{\mu\nu}_i}_{;\mu} = 0 \quad ,
\end{equation}
leading to the expressions
\begin{equation}
\rho_m = \frac{\rho_{m0}}{a^3} \quad , \quad
\rho_r = \frac{\rho_{r0}}{a^4} \quad , \quad
\rho_c = \sqrt{A + \frac{B}{a^6}} \quad , \quad \rho_\Lambda = \mbox{constant} \quad .
\end{equation}
Henceforth, we will fix the value of the scale factor today as equal to one, $a_0 = 1$. In this case,
$\rho_{m0}$, $\rho_{r0}$ and $\rho_{c0} = \sqrt{A + B}$ are the density for pressurelles matter, radiation and
the Chaplygin gas today. In the expression for the Chaplygin gas, the integration constant $B$ can be expressed
in terms of $A$ and $\rho_{c0}$, leading to $B = \rho_{c0}^2 - A$, what allows to rewritte the expression for
the Chaplygin gas density as
\begin{equation}
\label{chap-dens}
\rho_c = \rho_{c0}\sqrt{\bar A + \frac{1 - \bar A}{a^6}} \quad , \quad \bar A = \frac{A}{\rho_{c0}^2} \quad .
\end{equation}
The redefined parameter $\bar A$ will be connected with the sound velocity for the Chaplygin gas today as we will see
later.
\par
Using the equations of motion and the previous result, the first and the second derivatives of the
scale factor can be expressed in terms of the mass parameters and the scale factor itself:
\begin{eqnarray}
\label{f1}
\dot a &=& F(a) = \biggr\{R\biggr[\frac{\Omega_{m0}}{a} + \frac{\Omega_{r0}}{a^2} + a^2\Omega_{\Lambda0} +
\Omega_{c0}\sqrt{a^4\bar A + \frac{1 - \bar A}{a^2}}\biggl]\biggl\}^{1/2} \quad , \\
\label{f2}
\ddot a &=& G(a) = \frac{R}{2}\biggr\{- \frac{\Omega_{m0}}{a^2} - 2\frac{\Omega_{r0}}{a^3} + \Omega_{c0}
\frac{2\bar Aa^3 - \frac{(1 - \bar A)}{a^3}}{\sqrt{\bar Aa^4 + \frac{1 - \bar A}{a^2}}} + 2\Omega_{\Lambda0} a\biggl\} \quad ,
\end{eqnarray}
where $\Omega_{r0}$, $\Omega_{m0}$, $\Omega_{c0}$ and $\Omega_{\Lambda0}$ are fractions of the radiation density,
pressurelles matter density, Chaplygin gas density and cosmological constant density with respect to the total
mass today, respectivelly. Of course, $\Omega_{r0} + \Omega_{m0} + \Omega_{c0} + \Omega_{\Lambda0} = 1$, since we have
assumed a flat Universe. Moreover, $R = 8\pi Gt_0^2\rho_0/3$, where $t_0$ is the present age of the Universe and
$\rho_0$ is the total density of the Universe. In this way, (\ref{f1},\ref{f2}) have become dimensionless equations,
the dots meaning derivatives with respect to $t/t_0$, that is, fractions of the age of the Universe or, if one
prefers, the redshift $z$.
\par
In general, expressions (\ref{f1},\ref{f2}) do not admit simple closed solutions, except for the case
where $\Omega_{c0} = \Omega_{\Lambda0} = 0$. However, since in more general cases numerical integration
will be needed, we will not consider that particular case, which, in any way, leads to background solutions
that do not allow simple analytical integration of the perturbed equations. However, expressions
(\ref{f1},\ref{f2}) will be crucial in settling out the perturbed equations which will be used in
the numerical integration procedure.
\par
Before to treat perturbatively the system described before, the cosmological parameters for each model will be specified.
First of all, the Hubble parameter today will be taken as
\begin{equation}
H_0 = 72\;\frac{km}{Mpc.s} \quad .
\end{equation}
This leads to an age of the Universe of the order
\begin{equation}
t_0 \sim 4.1\times10^{17}\; s \sim 13.3\;Gy \quad .
\end{equation}
This value is quite favored by the recent estimations of the anisotropy of cosmic microwave background,
supernova type Ia observations, being also in good agreement with recent re-estimations of the
age of galactic objects obtained through the Hipparcos program \cite{hipparcos1,hipparcos2}.
Using the Einstein's equation, the resulting total density of the Universe
is $\rho_0 \sim 10^{-29}\;g/cm^3$.
In the next section, we will analyze three main models:
\begin{itemize}
\item Model I: $\Omega_{c0} = \Omega_{\Lambda0} = 0$, which will be called {\it baryonic model};
\item Model II: $\Omega_{\Lambda0} = 0$, which will be called {\it Chaplygin gas model};
\item Model III: $\Omega_{c0} = 0$, which will be called {\it cosmological constant
or $\Lambda$CDM model}.
\end{itemize}
Using estimations of cosmological parameters, coming from the various observational programs in course, as
summarized in \cite{charles}, we will fix the following values for each model:
\begin{itemize}
\item Model I: $\Omega_{r0} = 10^{-4}$; $\Omega_{m0} = 0.9999$;
\item Model II: $\Omega_{r0} = 10^{-4}$; $\Omega_{m0} = 0.2999$; $\Omega_{c0} = 0.7$;
\item Model III:  $\Omega_{r0} = 10^{-4}$; $\Omega_{m0} = 0.2999$; $\Omega_{\Lambda} = 0.7$.
\end{itemize}
The values used above are not the unique possibility of course. But, they are consistent with 
the different observational estimations: The data used above are well inside the observed values,
taking into account the bar errors.
For the estimation of the radiative contribution, neutrino and photon components are taken
into account. For the pressurelless fluid, baryonic and weakly interacting massive particles (WIMPS), which
must be the components of the cold dark matter,
are considered. In model II, specifically, the possible contribution of the cosmological constant
is replaced by the Chaplygin gas, while in model I both exotic fluids are ignored, only radiation and
pressurelless matter being kept.
\par
Each of the models described before imply a different moment for the equilibrium between radiation and
the other forms of matter. For a pure cosmological constant model, this equilibrium occurs for
$z_{eq} \sim 3,000$. For the Chaplygin gas models, this moment depends on the value of $\bar A$:
For $\bar A = 0.5$, it occurs at $z_{eq} \sim 4,600$, for $\bar A = 0.8$ at $z_{eq} \sim 6,100$, and
for $\bar A = 0.95$ at $z_{eq} \sim 8,000$. The equilibrium between radiation and matter is reached
at $z_{eq} \sim 10,000$ for the baryonic model. These different values for $z_{eq}$ do not play a
significant role since the background models imply a smooth transition from the radiation dominated
era to the matter dominated era, the transition extending for a non negligible period of time.
Hence, the exact moment where the radiation density equals the matter density is of minor importance
in this scenario.
\par
An important point must be noticed. In model I, there is an initial radiative phase, followed by
a material phase. The Universe remains always in a subluminal expansion. On the other hand, for
models II and III, there is a moment, quite near today, where the subluminal expansion gives place
to a superluminal expansion, that is, inflation. This moment can be easily determined putting $G(a) = 0$. For the cosmological constant model, with the values given above,
inflation begins to occur at $z \sim 0.67$.
For the Chaplygin gas model, this depends crucially on the value of $\bar A$: For $\bar A = 0.5$, it occurs
at $z \sim 0.01$; for $\bar A = 0.8$, it occurs at $z \sim 0.22$; for $\bar A = 0.95$, it occurs at $z \sim 0.43$.
The greater $\bar A$, the earlier the superluminal expansion of the Universe begins. This is natural, since
as $\bar A \rightarrow 0$, the Chaplygin model coincides with the baryonic model, where no superluminal expansion
occurs at any time; on the other hand, as $\bar A \rightarrow 1$, the Chaplygin gas model coincides with the
cosmological constant model.
\par
Finally, in what concerns specifically the Chaplygin gas model, there is an initial radiative phase, followed
by a matter dominated phase. The final stage is a deSitter phase, with the Chaplygin gas playing the role of
a cosmological constant of value $\rho_\Lambda = \sqrt{A}$. The Chaplygin gas behaves initially as a dust fluid,
but it implies an intermediate phase characterized by a mixing of stiff matter and cosmological constant \cite{pasquier}.
In \cite{bento}, a "generalized Chaplygin gas" was studied, considering an equation of state of the
type $p = -A/(\rho^\alpha)$, where that intermediate stage becomes a mix of cosmological constant and a fluid
characterized by $p = \alpha\rho$. In spite of being interesting to consider this more general equation
of state, we will restricted ourselves to the traditional Chaplygin gas, with an equation
of state given by (\ref{eq-est}).

\section{Perturbative analysis and the mass power spectrum}

In order to perform the perturbative analysis necessary to determine the mass power spectrum today, we
return back to the einsteinian equations, writing them as
\begin{equation}
R_{\mu\nu} = 8\pi G\sum_i^n (T^i_{\mu\nu} - \frac{1}{2}g_{\mu\nu}T^i) \quad , \quad 
{T^i_{\mu\nu}}^{;\mu} = 0 \quad ,
\end{equation}
where the indice $i$ labels the different matter components described in the preceding section.
The background solutions are described by the functions (\ref{f1},\ref{f2}).
The perturbations are introduced by writting the dynamical variables as
\begin{equation}
g_{\mu\nu} =\; \stackrel{0}{g}_{\mu\nu} + h_{\mu\nu} \quad , \quad \rho_i =\; \stackrel{0}{\rho}_i + \delta\rho_i \quad ,
\quad p_i =\;\stackrel{0}{p}_i + \delta p_i \quad ,
\end{equation}
where $\stackrel{0}{g}_{\mu\nu}$, $\stackrel{0}{\rho}_i$ and $\stackrel{0}{p}_i$ represent the background solutions
and $h_{\mu\nu}$, $\delta\rho_i$ and $\delta p_i$ are small fluctuations around them. Due to the coordinate
reparametrization freedom, we can impose a coordinate condition, which we choose to be the synchronous one:
$h_{\mu0} = 0$.
\par
We will consider just adiabatic perturbations. In this case,
the velocity of the sound for each component is given by
\begin{equation}
\frac{v_s^2}{c^2} = \frac{\partial p}{\partial\rho} \quad .
\end{equation}
For the Chaplygin gas, this reads
\begin{equation}
\frac{v_s^2}{c^2} = \frac{A}{\rho_c^2} \quad ,
\end{equation}
which is positive. Hence, the sound velocity of the Chaplygin gas is given today by
$v_s = \sqrt{\bar A}c$. As $\bar A$ approaches one, the sound velocity of the Chaplygin gas approaches
the velocity of light and, by (\ref{chap-dens}), the Chaplygin gas becomes essentially a cosmological constant.
\par
After a long but standard calculation \cite{weinberg,turner,brandenberg}, we end up with the following system of coupled
perturbed equations for the general model composed of radiation, pressurelless fluid, Chaplygin gas and
cosmological constant:
\begin{eqnarray}
\label{pe1}
\ddot h + 2\frac{\dot a}{a}\dot h &=& 3\biggr(\frac{\dot a}{a}\biggl)^2\biggr\{2\Delta_r\Omega_r +
\Delta_m\Omega_m + \biggr(1 + 3\frac{A}{\rho_c^2}\biggl)\Delta_c\Omega_c\biggl\} \quad ,\\
\label{pe2}
\dot\Delta_r + \frac{4}{3}\biggr(\Theta_r - \frac{\dot h}{2}\biggl) &=& 0 \quad , \\
\label{pe3}
\dot\Theta_r + \frac{\dot a}{a}\Theta_r &=& \frac{k^2}{4}\frac{\Delta_r}{a^2} \quad , \\
\label{pe4}
\dot\Delta_m &=& \frac{\dot h}{2} \quad , \\
\label{pe5}
\dot\Delta_c + \biggr[1 - \frac{A}{\rho_c^2}\biggl]\biggr(\Theta_c - \frac{\dot h}{2}\biggl) &=& -6\frac{\dot a}{a}
\frac{A}{\rho_c^2}\Delta_c \quad , \\
\label{pe6}
\dot\Theta_c + \biggr(2 + 3\frac{A}{\rho_c^2}\biggl)\frac{\dot a}{a}\Theta_c &=& k^2\frac{A}{\rho_c^2 - A}\frac{\Delta_c}{a^2} \quad .
\end{eqnarray}
In these equations $h = \frac{h_{kk}}{a^2}$, $\Delta_i = \frac{\delta\rho_i}{\rho_i}$ is the density contrast for each
fluid component, $\Omega_i = \frac{\rho_i}{\rho_T}$ denotes the mass fraction for each fluid at a given time, $\rho_T$ being
the total mass density at that time, $\Theta_i$ is the perturbed velocity for each fluid. A Fourier decomposition
has also been performed, writting all perturbed quantities as $\delta f(\vec x,t) = \delta f(t)e^{-i\vec k.\vec x}$,
$\vec k$ being the wavevector associated with each Fourier mode. Notice that the fluid velocity perturbation for
the pressurelless fluid and the density perturbation of the cosmological constant are zero. In particular, the
cosmological constant influences the perturbed equations through the behaviour of the background quantities only.
\par
Since we do not have in the general case an analytical expression for the scale factor as function of the time,
we rewritte the above equations in terms of the scale factor itself, using (\ref{f1},{\ref{f2}). Moreover, we
make equations (\ref{pe1},\ref{pe2},\ref{pe3},\ref{pe4},\ref{pe5},\ref{pe6}) dimensionless by multiplying them conveniently by
$t_0$, the age of the Universe, and reinserting, when is the case, $c$.
Finally, we eliminate $h$, expressing it in terms of $\Delta_m$, using (\ref{pe4}). The perturbed equations
read then
\begin{eqnarray}
\label{fpe1}
\Delta_m'' + \biggr\{\frac{G(a)}{F(a)} + \frac{2}{a}\biggl\}\Delta_m' &=& \frac{3}{2a^2}\biggr\{2\Delta_r\Omega_r +
\Delta_m\Omega_m + (1 + 3E(a))\Delta_c\Omega_c\biggl\} \quad ,\\
\label{fpe2}
\Delta_r' + \frac{4}{3}\biggr\{\frac{\Theta_r}{F(a)} - \Delta_m'\biggl\} &=& 0 \quad , \\
\label{fpe3}
\Theta_r' + \frac{\Theta_r}{a} &=& \biggr(\frac{80\pi k}{k_0}\biggl)^2\frac{\Delta_r}{4a^2F(a)} \quad , \\
\label{fpe4}
\Delta_c' + \biggr[1 - E(a)\biggl]\biggr\{\frac{\Theta_c}{F(a)} - \Delta_m'\biggl\} &=& - 6\frac{E(a)}{a}\Delta_c \quad ,\\
\label{fpe5}
\Theta_c' + \biggr[2 + 3E(a)\biggl]\frac{\Theta_c}{a} &=& \biggr(\frac{80\pi k}{k_0}\biggl)^2\frac{E(a)}{a^2F(a)(1 - E(a))}\Delta_c \quad ,
\end{eqnarray}
where the primes mean derivative with respect to $a$ and $E(a) = \bar Aa^6/[\bar Aa^6 + (1 - \bar A)]$.
The mass parameter are written as
\begin{equation}
\Omega_r = \frac{\Omega_{r0}}{F^2(a)a^2} \quad , \quad \Omega_m = \frac{\Omega_{m0}}{F^2(a)a}
\quad , \quad \Omega_c = \Omega_{c0}\sqrt{\bar A + \frac{1 - \bar A}{a^6}}\frac{a^2}{F^2(a)}
\quad .
\end{equation}
The wavenumber has been conveniently reparametrized in terms of a reference scale $k_0 = 2\pi/\lambda_0$,
where $\lambda_0 = 100\;Mpc$. Notice that, since we have imposed the value of the scale factor today equal
to one, the co-moving wavelengths are equal to the physical wavelengths today.
\par
Equations (\ref{fpe1},\ref{fpe2},\ref{fpe3},\ref{fpe4},\ref{fpe5}) represent the complete equations
for the Chaplygin gas model, with $\Omega_{\Lambda0} = 0$ in (\ref{f1},\ref{f2}). The cosmological constant
and the baryonic models are obtained by fixing $\Delta_c = \Theta_c = 0$ in the perturbed equations above.
For the cosmological constant model, it must be fixed
$\Omega_{c0} = 0$ in the unperturbed equations, while for the baryonic model it must also be fixed $\Omega_{\Lambda0} = 0$.
\par
The perturbed equations written above are completely consistent mathematically with the background model described
by (\ref{f1},\ref{f2}). However, from the physical point of view, there are many simplifications which must
be pointed out. First, what is called here "baryonic" matter is a general word for "pressurelles fluid", and
it includes both the baryonic matter properly speaking and the cold dark component (WIMPS). Even if the cold dark
matter component can also be represented by a zero pressure fluid at the stage of galaxy formation, it decouples from
the photonic fluid earlier than the baryonic matter due to its very small cross section. This fact has consequences for
the mass power spectrum; but we neglect it here, since it will affect the final results in the same way for
the three models to be studied.
\par
The second important approximation concerns the neutrino free streaming. Of course, the density of neutrinos
will affect the evolution of the background. However, due to its decoupling at quite early times, the neutrinos will
not contribute to the radiation pressure as the photonic component does in the gravitational collapse
process. Hence, in principle, this should be taken
into account in the perturbative equations. Again, this effect will be neglected: The radiation density in
the perturbed equations will be taken as in the background equations. As in the preceding case, since
this effect will be neglected in the same way in all three models to be studied, this approximation does not
spoil significantly the results from the point of view of identifying the main features of the Chaplygin gas
compared with the pure baryonic Universe and, mainly, with the cosmological constant. Another simplification concerns
the photon diffusion, which will not be considered also, since it leads to the breakdown of the perfect fluid
approximation.
\par
Having established the main features of the perturbative analysis to be carried out, we turn to some important
aspects of the problem. Equations (\ref{fpe1},\ref{fpe2},\ref{fpe3},\ref{fpe4},\ref{fpe5}) are complicated enough
to prevent us to try to obtain analytical solutions, even in the long wavelength limit. Hence, we will integrate them numerically. This integration
is carried out from $z \sim 10^{4}$ (approximatelly the moment of equilibrium between matter and radiation) until
$z = 0$ (today). It remains the important question of the initial conditions. The initial velocity perburbation 
$\Theta_i$ will be set equal to zero at that moment, as well as the derivative of the perturbations of the matter components,
$\Delta_i' = 0$. For the matter perturbations themselves, the most natural choice, and the more consistent
with respect to the results of the anisotropy of the cosmic microwave background, is the Harrison-Zeldovich spectrum,
which states that all perturbations have the same amplitude when they cross the horizon.
The Harrison-Zeldovich spectrum implies that \cite{turner}
\begin{equation}
\label{h-z}
(\delta_k)_H \propto k^{-3/2} \quad .
\end{equation}
However, it comes out to be more convenient to impose the initial conditions at a constant time $t_i$, instead
to impose them at the horizon crossing moment, since this crossing occurs at different moments for different wavelengths.
In order to obtain a spectrum near the Harrison-Zeldovich spectrum at the horizon crossing, the spectrum at a constant
time defined before this crossing must be different from (\ref{h-z}). In \cite{turner} two
phases, the radiative and material ones, were considered and
it came out that the spectrum at constant time should be written as
\begin{equation}
\label{c-t-s}
(\delta_k)_t \propto k^{1/2} \quad ,
\end{equation}
in order to obtain a Harrison-Zeldovich spectrum at horizon crossing.
In our model, the situation is more complex since there are not two so well definite phases.
However, (\ref{c-t-s}) remains a reasonable approximation, as it has been verified
numerically. It must be stressed, in any case, that the goal of the computation performed
in this work is not to test the primordial spectrum, but to identify the specific features
of the Chaplygin gas. In figures $1$, $2$ and $3$ the behaviour of the density contrast
for matter, Chaplygin gas and radiation are shown for the Chaplygin gas model with
$\bar A = 0.8$ and $k/k_0 = 0.5$.
\par
We have evaluated numerically the power spectrum today for the dust component, represented
by the baryonic and cold dark matter component, given their power spectrum at $t = t_i$ (equivalent to
$z = 10^4$). We restrict ourselves to these components since they are essentially what is observed
in clusters and super-clusters of galaxies. In figure $4$ we display the results for $P(k) = k^{3/2}\delta_k$ as function of $k$. The numerical
integration was performed from scales of the order of the Hubble horizon today, $\lambda_{max} \sim d_H \sim 4,000\;Mpc$ for the value of the
Hubble parameter chosen, down to $\lambda_{min} = 100\;Mpc$. The normalization of each spectrum is made by its value
at $\lambda_{max}$. Hence, each curve in figure $4$ has a slightly different parametrization. However, what is
more relevant is the slope of each curve. In this aspect we notice that all three models
reveal an increasing spectrum. The slope of the spectrum is higher
for the baryon model, becoming smaller for the Chaplygin gas model, approaching the cosmological constant model
as $\bar A \rightarrow 1$. Hence, the Chaplygin gas interpolates a pure baryonic model and a $\Lambda$CDM
model as the velocity of the sound associated with it goes from zero to the velocity of light.
In figure $5$ we plot the same spectrum for $\ln P^2(k) = \ln(k^{3}\delta_k^2)$ as function of $\ln\lambda$. Notice,
from this figure, that the spectrum for the cosmological constant tends to be more flat, what is compatible
with the CMB measurements \cite{charles}, deviating significantly from this "almost Harrison-Zeldovich" spectrum
as we approach the pure baryonic spectrum, making the velocity of sound of the Chaplygin gas going from
$v_s^2 = c^2$ to $v_s^2 = 0$.
\par
The interpretation of these results makes appeal for some physical aspects of the problem. Notice that
the three models contain the same amount of radiative fluid. As the perturbation enters in the horizon,
the radiative fluid contributes to the pressure and may distort the initial spectrum. In principle, this
occurs in the same way for the three models. A quite significant effect may come from
the moment where the perturbation enters in the horizon, which is evidently different for each model. The condition for the horizon crossing is
given by the relation $k = aH/c$. Using the parametrizations described before, this
relation takes the form $k/k_0 = F(a)/(40)$. A perturbation with a scale of the order of
$100\;Mpc$ today has entered in the horizon at $z \sim 1,402$ for the baryonic model, while for the
cosmological constant model it has entered at $z \sim 2,771$. For a perturbation of
the order of $2,000\;Mpc$ today, the horizon crossing occured at $z \sim 12$ for the
cosmological constant model, and at $z \sim 3$ for the pure baryonic model.
The Chaplygin gas models leads to values between those
two. As the cosmological constant model is approached, the perturbations analyzed have entered earlier in the
horizon, being subjected for a longer time to the effects of pressure. 
\par
Another important effect, which seems to be more significant than the previous one, comes from the moment the Universe begins to accelerate. The expansion of the Universe
leads to a friction like term in the perturbative equations (see, for example, (\ref{pe1})). The accelerated expansion
tends to smooth the fluctutations more efficiently than a non-accelerated expansion. The baryonic model never
accelerates, as it was discussed in the preceding section. The cosmological constant model begins
to accelerate quite recently, but in such a way that the fluctuations in density are significantly affected.
Of course, the Chaplygin gas models leads to an accelerated Universe whose time transition is near to our present time if
$\bar A$ is small, and near to the cosmological constant model time transtion as $\bar A$ approaches one. This fact is reflected
in the final evaluated spectrum, in special in the slope of the spectrum.

\section{Conclusions}

The Chaplygin gas appears as an alternative to the cosmological constant and to quintessence in order
to explain the stage of accelerated expansion of the Universe today. It is a fluid with negative pressure
which depends on the inverse of the density. Due to its equation of state, it interpolates a matter dominated
phase and a cosmological constant phase in what concerns the evolution of the Universe. In this paper
we have addressed the question of structure formation in the realm of a Universe dominated by
the Chaplygin gas. The goal of the present work was to determine the features of the mass power spectrum
inside the horizon today given a primordial spectrum in a given time near the equilibrium between
radiation and matter, around $z \sim 10^4$. In order to track the specific behaviour of the Chaplygin gas,
we computed also the mass spectrum for a pure baryonic Universe and for a $\Lambda$CDM model.
\par
An important parameter in the Chaplygin gas model is played by the constant $A$ in the equation of
state of this fluid. It is related to the sound velocity in such a fluid. The sound velocity of the
Chaplygin gas today is given by $\bar A = \frac{A}{\rho_{c0^2}}$, where $\rho_{c0}$ is the Chaplygin gas
density
today. If $\bar A = 0$, the Chaplygin gas behaves always as a dust fluid; as $\bar A$ approaches one,
the velocity of the sound approaches the velocity of light, and the Chaplygin gas become indistinguishable
from a pure cosmological constant.
\par
This fact is reflected in the mass power spectrum. We have evaluated this spectrum numerically, using
an almost initial Harrison-Zeldovich spectrum. Considering scales between the Hubble horizon and $100\;Mpc$
(the normal threshold scale above which observations indicate that
the Universe may be considered essentially homogeneous and isotropic),
we find that, for the models studied here, the spectrum increases with the wavenumber $k$. However,
the slope is higher for a pure baryonic Universe, and smaller for a $\Lambda$CDM model. The Chaplygin
gas model interpolates between those two as $\bar A$ goes from zero to one. This smoothing of the spectrum
can be understood by remembering that the expansion of the Universe plays the role of a friction term,
and an accelerated expansion tends to suppress more efficiently the local irregularities than a non-accelerated
expansion. For the chosen values of the cosmological parameters, in special the mass parameters, the
more one approaches a cosmological constant model, the earlier the Universe begins to accelerate,
and consequently the power spectrum becomes more flat. We notice, moreover, that the earlier the Universe begins to
accelerate, the earlier the perturbation of a given scale enters in the horizon, being subjected no longer
only to the the expansion but also to pressure.
\par
A next step would be to compare the results obtained for the spectrum with the observational data.
For the scales chosen, this means that high redshift surveys of structures must be compared with
the theoretical models. Important programs in this sense are in course \cite{lahav,peacock,percival}.
The data concerning the mass power spectrum are resctricted mainly
to scales up to $\lambda \sim 800\;Mpc$ \cite{lahav}, but the
uncertainty is higher the higher is the redshift. This observational data cover only the higher values
of the wavenumber in the spectrum we
have determined here. But, we think that no detailed comparison, even in this interval, of our results with
the observational data may be possible until a more realistic perturbative analyis is performed,
mainly in what concerns to take into account the specific features of the cold dark matter WIMPS and
the neutrino free streaming. In very small scales, non-linear effects must also be taken into account. Notice that, in spite of all the simplifications, the qualitative features of
the spectrum found are similar to the observational results (see \cite{peacock}, for example, and
also \cite{peacock1}).
In this sense, the present work is just a first step in the investigation
if a Chaplygin gas model can be competitive with other models, like $\Lambda$CDM or quintessence.

\vspace{0.5cm}
\noindent
{\bf Acknowledgements:}  We thank Fl\'avio Gimenes Alvarenga and Antonio Brasil Batista for
the careful reading of the manuscript. This work has receveid partial financial supporting from CNPq (Brazil).

\newpage
\centerline{\bf Figures captions}

\vspace{2.cm}
\begin{itemize}
\item Figure 1: Behaviour of the density contrast for the ordinary matter, with $\bar A = 0.5$ and $k/k_0 = 0.5$.
\item Figure 2: Behaviour of the density contrast for the Chaplygin gas, with $\bar A = 0.5$ and $k/k_0 = 0.5$.
\item Figure 3: Behaviour of the density contrast for radiation, with $\bar A = 0.5$ and $k/k_0 = 0.5$.
\item Figure 4: The power spectrum $P(k) = k^{3/2}\delta_k$ as function of $k/k_0$.
\item Figure 5: Behaviour of $\ln P^2(k)$ as function of $\ln\lambda/\lambda_0$.
\end{itemize}
\end{document}